\documentclass[aps,letterpaper,twocolumn,preprintnumbers,floatfix,superscriptaddress]{revtex4}

\usepackage{graphicx}
\usepackage{dcolumn}
\usepackage{bm}
\usepackage{epsfig}
\usepackage{gensymb}
\usepackage{mathrsfs}
\usepackage{amsmath}
\usepackage{amssymb}
\usepackage{graphicx,bm}
\usepackage{natbib}
\usepackage{slashed}
\usepackage[makeroom]{cancel}
\usepackage{dsfont}
\usepackage{longtable}
\usepackage{multirow}
 \usepackage[titletoc]{appendix}

\def\fun#1#2{\lower3.6pt\vbox{\baselineskip0pt\lineskip.9pt
  \ialign{$\mathsurround=0pt#1\hfil##\hfil$\crcr#2\crcr\sim\crcr}}}

\usepackage{tikz}
\usetikzlibrary{decorations.pathmorphing,decorations.markings,trees,shapes}

\def\lsim{\mathrel{\rlap{\raise 2.5pt \hbox{$<$}}\lower 2.5pt\hbox{$\sim$}}}
\def\gsim{\mathrel{\rlap{\raise 2.5pt \hbox{$>$}}\lower 2.5pt\hbox{$\sim$}}}

\def\f{kinematic factor}

\input epsf

\usepackage{color}

\newcommand{\comment}[1]{}

\newcommand{\be}{\begin{equation}}
\newcommand{\ee}{\end{equation}}
\newcommand{\bea}{\begin{eqnarray}}
\newcommand{\eea}{\end{eqnarray}}

\newcommand{\vev}[1]{\langle #1 \rangle}
\newcommand{\ket}[1]{| #1 \rangle}

\newcommand{\hc}{{\textit{h.c.}}~}

\newcommand{\eq}[1]{\begin{equation}\begin{split} #1 \end{split}\end{equation}}

\newcommand*\Dbox{\mathop{}\!\mathbin\Box}

\begin{document}

\title{Standard Model Effective Field Theory from On-shell Amplitudes }

\author{Teng Ma}
\affiliation{
CAS Key Laboratory of Theoretical Physics, Institute of Theoretical Physics,
Chinese Academy of Sciences, Beijing 100190, China.}
\author{Jing Shu}
\affiliation{
CAS Key Laboratory of Theoretical Physics, Institute of Theoretical Physics,
Chinese Academy of Sciences, Beijing 100190, China.}
\affiliation{School of Physical Sciences, University of Chinese Academy of Sciences, Beijing 100049, P. R. China.}
\affiliation{CAS Center for Excellence in Particle Physics, Beijing 100049, China}
\affiliation{Center for High Energy Physics, Peking University, Beijing 100871, China}
\author{Ming-Lei Xiao}
\affiliation{
CAS Key Laboratory of Theoretical Physics, Institute of Theoretical Physics,
Chinese Academy of Sciences, Beijing 100190, China.}

\begin{abstract}

We present a general method of constructing unfactorizable on-shell amplitudes (amplitude basis), and build up their one-to-one correspondence to the independent and complete operator basis in effective field theory (EFT). We apply our method to the Standard Model EFT, and identify the amplitude basis in dimension 5 and 6, which correspond to the Weinberg operator and operators in Warsaw basis except for some linear combinations.

\end{abstract}

\pacs{xxx}

\maketitle

\section{Introduction}

Effective Field theory (EFT) has wide applications in various aspects of physics. It serves as a powerful tool to understand the emergent coarse-graining behavior where the underlying system has sophisticated patterns or is strongly coupled, such as superconductivity~\cite{Ginzburg:1950sr}, fractional quantum hall effect~\cite{AQHE}, or low energy QCD~\cite{Weinberg:1968de}, etc. Meanwhile, EFT provides a model-independent method to categorize and parametrize possible unknown physics from the ultraviolet (UV). In this case, our best example is the Standard Model (SM) EFT, which becomes a basic paradigm of exploring the imprints of beyond the SM effects.

One essential step in an EFT calculation is the identification of complete operator basis~\cite{Buchmuller:1985jz, Grzadkowski:2010es,Lehman:2014jma,Liao:2016hru,Liao:2016qyd}. To construct the full set of independent operators in quantum field theory, one has to eliminate the redundancies from equation of motion (EOM) and integration by parts (IBP), which yield relations between operators. Previously, people rely on symmetry~\cite{Jenkins:2009dy,Lehman:2015via,Lehman:2015coa,Henning:2015alf,Henning:2017fpj,Gripaios:2018zrz} in SM to eliminate those redundancies and encode enumeration of operators in a Hilbert series. Nevertheless, those methods do not give exact expressions of all operators and are not naturally directed to the SMEFT calculations.

In this paper, we introduce a novel way to write down all independent operators, which is based on the on-shell amplitude method~\cite{Witten:2003nn}. Instead of using symmetries to deal with EOM and IBP, we can write down all complete local on-shell amplitudes respecting Lorentz symmetry, SM gauge symmetry and spin-statistics theorem. Those on-shell amplitudes are in one to one correspondence with the operators, which naturally form a new \emph{amplitude basis}. Our key observation is that for amplitude basis, the elimination of EOM and IBP redundancies are trivially realized by external leg on-shell conditions and momentum conservation, as naturally inherited from the on-shell amplitudes. This approach was used recently to infer the EFT Lagrangian for theories with a spin-0 or 1 singlet coupled to gluons~\cite{Shadmi:2018xan}.

Our method really has its own advantages to begin with. By using the on-shell amplitude method, the root of our amplitude basis is nothing but the unfactorizable on-shell amplitudes from locality (positive power of Madelstam variables without poles). We start directly with the computation of unfactorizable amplitudes from spin helicity formulism, thus automatically provide the basic building blocks of EFT calculation, and  advanced techniques like recursion relations, unitarity cuts, etc can be used naturally. Indeed, our method also greatly simplifies the calculation. The amplitude basis in d=6 for SMEFT corresponds to the Warsaw basis, except for some linear combinations, can be derived in few pages later in the paper.

The rest of the paper is organized as follows. We first discuss the general structure of the amplitude basis and outline the rules of constructing the complete set of amplitude basis in a given dimension. Armed with those tools, we explicitly construct the amplitude basis for SMEFT in $d=5$ and $6$, and map them to the corresponding operators. Finally we conclude.

\section{The Structure of Amplitude Basis}
\label{sec:Amp_Basis}
We start from $S$-matrix program, which uses a set of low point amplitudes as building blocks, and construct higher point amplitudes by matching their residues using recursion relations. Because of the renormalizability, the finite set of low point amplitudes is sufficient as the theory input. However, in an non-renormalizable theory like SMEFT, the irrelevant operators $\mathcal{O}$ are independent interactions, which cannot be on-shell constructed by recursion relations without the help of symmetries. Therefore, those independent amplitudes should be viewed as the input basis of the theory, which correspond to the independent set of operators and can be classified by their dimensions.

We build the one-to-one correspondence between the amplitude basis and irrelevant operators $\mathcal{O}$ by enforcing that all fields from $\mathcal{O}$ are on shell. The gauge symmetry, which reflects the redundancy, can be used to reduce the independent amplitudes. Only the \textit{leading} contact on-shell amplitudes which have the minimal fields from $\mathcal{O}$ are enough to fully construct all on-shell amplitudes for a given dimension~\cite{Cohen:2010mi}. Indeed, this is the same as the famous example of Yang-Mills theory, the cubic term $A^2 \partial A$ captures the full information for the on-shell amplitudes,
and the 4-point on-shell amplitudes are not independent from recursion relations (the existence of 4-point contact interactions from $A^4$ simply has no on-shell information).  Since the above arguments only exploit gauge invariance, they should apply to the non-renormalizable theory.

From Lorentz symmetry, the basic building blocks to construct the effective operators are $F^\pm_{\mu \nu} \equiv \frac{1}{2}(F_{\mu \nu} \pm i \tilde{F}_{\mu \nu} )$ ($\tilde{F}_{\mu \nu} \equiv \epsilon_{\mu \nu \rho \sigma} F^{\rho \sigma}$), $\psi_L$, $\psi^c_L$, $\phi$ and covariant derivative $D_\mu$ which transforms under Lorentz group $SU(2)_L \times SU(2)_R \equiv SO(3,1)$ as $(1,0)$, $(0,1)$, $(1/2,0)$, $(0,1/2)$ and $(0,0)$. As we have mentioned above, in the amplitude operator correspondence, only the leading contact on-shell amplitudes are taken into account, this suggests that we take ``$F_{\mu \nu} \rightarrow \partial_\mu A_\nu - \partial_\nu A_\mu$" and ``$D_\mu \rightarrow \partial_\mu$" to construct those effective operators.

There are two redundancies in the effective operators from Equation of Motion (EOM) and Integration by Part (IBP). However, in our amplitude basis, it is taken care automatically because of the on shell condition and momentum conservation. EOM for each fields will transmit the operators involving the derivatives of such fields into other operators. In our definition, the on shell conditions $p^2=0$ suggests that operators involving $\Dbox\phi$, $D\!\!\!\!/\psi$ or $D_{\mu}F^{\mu\nu}$ should vanish in the amplitude basis, thus there is no such redundancy. For IBP in the amplitude basis, two operators differing by a total derivative are equivalent as the total derivative is the sum of all external momentum which equals zero.

The general on-shell scattering amplitude should have the following form
\eq{\label{eq:amp_basis_ver1}
	\mathcal{M}_{\{\alpha\}} = f(\lambda_i,\tilde\lambda_i) g(s_{ij}) T_{\{\alpha\}},
}
where $\lambda_i,\tilde\lambda_i$ are helicity spinors of the $i$th leg and Mandelstam variable $s_{ij} \equiv 2p_i.p_j$ (see more details about spinor formalism in App.~\ref{sub:spinor}). $f$ is the little group weight function which is a function of spinor products $[ij]=\lambda_i\epsilon\lambda_j$ and $\vev{ij}=\tilde\lambda_i\epsilon\tilde\lambda_j$ and $g$ is the little group invariant function. $T$ is the group factor, bearing all the internal group indices of the external legs $\{\alpha\}$ and forming invariant tensors. For amplitude basis in a non-renormalizable theory, all spinor products in $f$ and $g$ have positive powers, which have no physical poles in Mandelstam variables and can not be factorized into smaller building blocks due to locality~\cite{Elvang:2010jv}.

According to dimension counting, it's easy to get the operator dimension $d$ of an amplitude basis
\eq{\label{eq:dim_ct}
	d = n+m =n + [f]+[g],
}
where $n$ is the number of legs and $[f]$ and $[g]$ are the dimensions of $f$ and $g$  where $[g]$ is always an even integer. The scattering processes are classified in terms of fermion number $n_\psi$ and gauge boson number $n_A$. Each fermion contributes one helicity spinor, and each gauge boson contributes two, thus the number of spinor products is at least $m \geq \frac{1}{2}n_\psi+n_A$. Using Eq\eqref{eq:dim_ct} we have
\eq{
	\frac{3}{2}n_\psi + 2 n_A \leq d
}
which gives a finite possibilities below a certain dimension. For instance, to get amplitude basis below $d=6$, what we need are scattering amplitudes with $(n_\psi,n_A)$ satisfying $\frac{3}{2}n_\psi + 2n_A \leq 6$. We list all possible amplitude basis in $(n_\psi,n_A, h)$ where $h$ is the total helicity with $h \ge 0$ in Table~\ref{tab:Lfactor} since we can just flip the helicity for $h<0$. We leave scalar number unspecified to shorten the list, because adding scalar does not change the form of Lorentz factor. For each $(n_\psi, n_A, h)$, we examine all possible helicity combinations (up to the conjugation).

For a given helicity assignment, we can write down the net powers of helicity spinors of all legs. For instance
\eq{\label{eq:f_ex}
	& f(\psi^+\psi^+\phi^2) \sim \lambda_1\lambda_2, \\
	& f(\psi^+\psi^-\phi^2) \sim \lambda_1\tilde\lambda_2.
}
To contract the spinor indices, we use the complete set of Clifford algebra $\{1,\sigma^{\mu},\sigma^{\mu\nu},\sigma^{\mu\nu\rho},i\epsilon^{\mu\nu\rho\xi} \}$ to construct bilinears. Specifically, $\{1,\sigma^{\mu\nu}\}$ can be used to contract two $\lambda$s or two $\tilde{\lambda}$s, while $\{\sigma^{\mu},\sigma^{\mu\nu\rho}\}$ can be used to contract a $\lambda$ and a $\tilde\lambda$. The more spacetime indices ($\mu,\nu,\dots$) a bilinear has, the more momenta we need to add to contract with them, which increases $m$. For instance to contract a $\lambda_i$ and a $\tilde\lambda_j$, the lowest dimension combination we can write down is $(\lambda_i\sigma^{\mu}\tilde{\lambda}_j) p_{k\mu} \equiv [i|p_k\ket{j} = [ik]\vev{kj}$. Following this rule, the lowest dimension amplitudes for the cases \eqref{eq:f_ex} are
\eq{
	& f(\psi^+\psi^+\phi^2) \sim (\lambda_1 \epsilon \lambda_2) \equiv [12], \\
	& f(\psi^+\psi^-\phi^2) \sim (\lambda_1 \epsilon \sigma^{\mu}\tilde\lambda_2)p_{3 \mu} \equiv [1|p_3 \ket{2}.
}
Moreover, Mandelstam variables can be added freely to $g$, as it is helicity blind. For each helicity assignment, there is a \f ~with minimum $m$, and thus a minimum dimension $d$, which we call \emph{primary amplitude}. It is the leading amplitude for a given scattering states.

		\begin{table}
			\begin{tabular}{|c|c|c|c|c|}
				\hline
				$(n_\psi,n_A,h)$ &  Primary amplitude & $m_{min}$ & $n_s$ & $d_{min}$ \\
				\hline
				(0,0,0)		&	$f(\phi^{n_s})=1$ & 0 &$n_s \geqslant 3$ & 3				\\
		
				\hline
				(0,2,2)		&	$f(A^+A^+\phi^{n_s}) = [12]^2$ & 2	& & 5		 \\
					\hline
				(0,3,3)		&	$f(A^+A^+A^+) = [12][23][31]$  & 3	& & 6			 \\
				\hline
				(2,0,1)		&	$f(\psi^+\psi^+\phi^{n_s}) = [12]$ & 1	& & 4				 \\
                \hline
                (2,0,0)     & $f(\psi^+\psi^-\phi^{2}) = [1|p_3\ket{2} $ & 2 & $n_s \geqslant 2$ & 6 \\

                \hline
				(2,1,2)		&	$f(A^+\psi^+\psi^+ \phi^{n_s}) = [12][13]  $ & 2	 & & 5			\\
				\hline
				(4,0,2)		&	$f(\psi^+\psi^+\psi^+\psi^+) = [12][34]^*$ & 2 & & 6		 \\
				\hline
				(4,0,0)			&	$f(\psi^+\psi^+\psi^-\psi^-) = [12]\vev{34}$ & 2 & & 6	 \\
				\hline
			\end{tabular}
			\caption{All classes of amplitude basis with $d \le 6$. The $*$ for the $(4,0,2)$ case stands for multiple ways of spinor contraction. }
			\label{tab:Lfactor}
		\end{table}

We work out all primary amplitudes in the TABLE~\ref{tab:Lfactor} for $d \le 6$. For the $(4,0,2)$ case, there are different ways of spinor contraction combinations, and we can define $f^{\pm}(\psi^+\psi^+\psi^+\psi^+) = ([13][24] \pm [14][23])$ after applying Schouten identity $[12][34] + [13][42] + [14][23] = 0$.

\section{Counting Effective Operators in SM EFT}

The results in the previous section are simple consequences of Lorentz invariance, gauge invariance and locality. When applying to SM matter fields in TABLE~\ref{tab:SM}, one has to take into account the SM gauge quantum numbers and respect Fermi or Boson statistics for identical fields. In this section, we explicitly construct the complete amplitude basis at dim-5 and dim-$6$ for SM EFT.

To count dim-5 amplitude basis in SM EFT, we need to combine the amplitudes in TABLE~\ref{tab:Lfactor} and appropriate group factors.
Group factors are not always unique for a given set of group indices, and when there are multiple choices, we use superscripts to label them. In particular, superscripts $\pm$ indicate permutation symmetry among the same type of indices, such as $T^{\pm}_{\alpha\beta\dot\alpha\dot\beta} = \frac{1}{2}(\delta_{\alpha\dot\alpha}\delta_{\beta\dot\beta} \pm \delta_{\alpha\dot\beta}\delta_{\beta\dot\alpha})$.
Among the \f s\ in TABLE~\ref{tab:Lfactor}, we find that only the $f(\psi^+\psi^+\phi^2)$ combination is the SM gauge singelt, which is
\eq{
	\mathcal{M}(L_{\alpha}L_{\beta}H_{\gamma}H_{\delta}) = 
	[12] (\epsilon_{\alpha\gamma}\epsilon_{\beta\delta} + \epsilon_{\alpha\delta}\epsilon_{\beta\gamma}) ,
}
where the group factor is chosen to satisfy the spin statistics. Together with its conjugate $f(\psi^-\psi^-\phi^2)$ with opposite helicity, we find the only 2 dim-5 amplitude basis in SM EFT, which are the Weinberg operators $\mathcal{O}^{(5)} = \frac{1}{\Lambda}(H L)^2 + \hc$.

Dim-6 amplitude basis in SM EFT are obtained in the same way. It is interesting that the classes of our amplitude basis in TABLE~\ref{tab:Lfactor} already reproduce the classes of operator summarized in~\citep{Grzadkowski:2010es} as the Warsaw basis. Because we choose the group factor basis and Mandelstam variables according to the permutation symmetry, the resultant amplitude basis for the same scattering states we get might be the linear combination of the operators defined in Warsaw basis. We list the correspondence below:
\begin{enumerate}
	\item Class $\mathcal{M}(\phi^{n_s})$ ($\mathcal{O}\sim\varphi^6$ and $\varphi^4D^2$):
	\begin{center}
	\begin{tabular}{|l|l|}
		\hline
		Operator 			&	Amplitude Basis 	\\
		\hline
		$\mathcal{O}_H$	&	$\mathcal{M}(H^3_{\alpha\beta\gamma}H^{\dagger3}_{\dot\alpha\dot\beta\dot\gamma}) = T^+_{\alpha\beta\gamma\dot\alpha\dot\beta\dot\gamma}$ \\
		\hline
		$2\mathcal{O}_{HD}-\mathcal{O}_{H\Dbox}$	&	$\mathcal{M}^+(H^2_{\alpha\beta}H^{\dagger2}_{\dot\alpha\dot\beta}) = s_{12}T^+_{\alpha\beta\dot\alpha\dot\beta}$ \\
		\hline
		$2\mathcal{O}_{HD}+\mathcal{O}_{H\Dbox}$	&	$\mathcal{M}^-(H^2_{\alpha\beta}H^{\dagger2}_{\dot\alpha\dot\beta}) = (s_{13} - s_{23})T^-_{\alpha\beta\dot\alpha\dot\beta}$ \\
		\hline
	\end{tabular}
	\end{center}
where  $T^+_{\alpha\beta\gamma\dot\alpha\dot\beta\dot\gamma} \equiv \delta_{\alpha \dot\alpha} \delta_{\beta \dot\beta}  \delta_{\gamma \dot \gamma} +\delta_{\beta \dot\alpha} \delta_{\alpha  \dot\beta}  \delta_{\gamma \dot \gamma}  + \delta_{\gamma \dot\alpha} \delta_{\beta \dot\beta}  \delta_{\alpha  \dot \gamma} + \delta_{\beta \dot\alpha} \delta_{\gamma \dot\beta}  \delta_{\alpha   \dot \gamma} +\delta_{\alpha \dot\alpha} \delta_{\gamma \dot\beta}  \delta_{\beta  \dot \gamma} +\delta_{\gamma \dot\alpha} \delta_{\alpha  \dot\beta}  \delta_{\beta  \dot \gamma} $ is fully symmetric for $SU(2)_L$ indices  $\alpha\beta\gamma$ and $\dot\alpha\dot\beta\dot\gamma$. $T^\pm _{\alpha\beta\dot\alpha\dot\beta} \equiv \delta_{\alpha \dot\alpha} \delta_{\beta \dot\beta}  \pm \delta_{\beta \dot\alpha} \delta_{\alpha \dot\beta} $ is the (ant-)symmetric group structure for indices $\alpha\beta$ and $\dot \alpha \dot \beta$. $s_{12}$ and $s_{12} - s_{23}$ are the symmetric and antisymmetric Mandelstam variables at the order $s$.

	\item Class $\mathcal{M}(A^+A^+\phi^2)$ and $\mathcal{M}(A^-A^-\phi^2)$ ($\mathcal{O}\sim X^2\varphi^2$):
	
	Instead of operators with definite CP, the amplitude basis are more naturally written down for definite chirality. They have easy linear relations.
	
	\begin{center}
	\begin{tabular}{|l|l|}
		\hline
		Warsaw 			&	Amplitude Basis 	\\
		\hline\hline
		$\mathcal{O}_{HB}+ \mathcal{O}_{H\tilde{B}}$	&	$\mathcal{M}(B^{+}B^{+}H_{\alpha}H^{\dagger}_{\dot\alpha}) = [12]^2\delta_{\alpha\dot\alpha} $ \\
		
     $\mathcal{O}_{HB}- \mathcal{O}_{H\tilde{B}}$         & $\mathcal{M}(B^{-}B^{-}H_{\alpha}H^{\dagger}_{\dot\alpha}) = \langle12\rangle^2\delta_{\alpha\dot\alpha}  $ \\
\hline
		$\mathcal{O}_{HWB}+\mathcal{O}_{H\tilde{W}B}$	&	$\mathcal{M}(B^{+}W^{i+}H_{\alpha}H^{\dagger}_{\dot\beta}) = [12]^2\tau^i_{\alpha\dot\beta}$ \\
$\mathcal{O}_{HWB}- \mathcal{O}_{H\tilde{W}B}$  & $\mathcal{M}(B^{-}W^{i-}H_{\alpha}H^{\dagger}_{\dot\beta}) = \langle12\rangle^2\tau^i_{\alpha\dot\beta}$     \\

		\hline
		$\mathcal{O}_{HW} + \mathcal{O}_{H\tilde{W}}$	&	$\mathcal{M}(W^{i+}W^{j+}H_{\alpha}H^{\dagger}_{\dot\beta}) = [12]^2 T^{ij+}_{\alpha\dot\beta}$ \\
$\mathcal{O}_{HW}- \mathcal{O}_{H\tilde{W}}$	&	$\mathcal{M}(W^{i-}W^{j-}H_{\alpha}H^{\dagger}_{\dot\beta}) =\langle12\rangle^2T^{ij+}_{\alpha\dot\beta}$ \\
		\hline
		$\mathcal{O}_{HG}+ \mathcal{O}_{H\tilde{G}}$	&	$\mathcal{M}(G^{A+}G^{B+}H_{\alpha}H^{\dagger}_{\dot\beta}) = [12]^2 T^{AB+}_{\alpha\dot\beta}$ \\
$\mathcal{O}_{HG}- \mathcal{O}_{H\tilde{G}}$	&	$\mathcal{M}(G^{A-}G^{B-}H_{\alpha}H^{\dagger}_{\dot\beta}) = \langle12\rangle^2T^{AB+}_{\alpha\dot\beta}$ \\
		\hline
	\end{tabular}
	\end{center}
where $\tau^i$ is the Pauli matrix, $T^{ij+}_{\alpha\dot\beta} \equiv \delta^{ij} \delta_{\alpha\dot\beta}$ and $T^{AB+}_{\alpha\dot\beta} \equiv \delta^{AB} \delta_{\alpha\dot\beta}$.

	\item Class $\mathcal{M}(A^+A^+A^+)$ and $\mathcal{M}(A^-A^-A^-)$ ($\mathcal{O}\sim X^3$):
	
	\begin{center}
	\begin{tabular}{|l|l|}
		\hline
		Warsaw 			&	Amplitude Basis 	\\
		\hline\hline
		$\mathcal{O}_W + \mathcal{O}_{\tilde{W}}$	&	$\mathcal{M}(W^{i+}W^{j+}W^{k+}) = [12][23][31]\epsilon^{ijk}$ \\
$\mathcal{O}_W - \mathcal{O}_{\tilde{W}}$	&	$\mathcal{M}(W^{i-}W^{j-}W^{k-}) = \langle12\rangle \langle23\rangle \langle31\rangle\epsilon^{ijk}$ \\
		\hline
		$\mathcal{O}_G + \mathcal{O}_{\tilde{G}}$	&	$\mathcal{M}(G^{A+}G^{B+}G^{C+}) = [12][23][31] f^{ABC}$ \\
        $\mathcal{O}_G - \mathcal{O}_{\tilde{G}}$	&	$\mathcal{M}(G^{A-}G^{B-}G^{C-}) =\langle12\rangle \langle23\rangle \langle31\rangle f^{ABC}$ \\
		\hline
	\end{tabular}
	\end{center}
where $\epsilon^{ijk}$ and $f^{ABC}$ are $SU(2)_L$ and $SU(3)_c$ structure constant.
	
	\item  Class $\mathcal{M}(\psi^+\psi^+\phi^3)$ ($\mathcal{O}\sim \psi^2\varphi^3$) + h.c.:
	
	\begin{center}
	\begin{tabular}{|l|l|}
		\hline
		Warsaw 			&	Amplitude Basis 	\\
		\hline\hline
		$\mathcal{O}_{eH}$	&	$\mathcal{M}(L_{\alpha}eH_{\beta}H^{\dagger2}_{\dot{\alpha}\dot{\beta}}) = [12]T^+_{\alpha\beta\dot\alpha\dot\beta}$ \\
		\hline
		$\mathcal{O}_{dH}$	&	$\mathcal{M}(Q_{a\alpha}d_{\dot{a}}H_{\beta}H^{\dagger2}_{\dot{\alpha}\dot{\beta}}) = [12]T^+_{\alpha\beta\dot\alpha\dot\beta}\delta_{a\dot{a}}$ \\
		\hline
		$\mathcal{O}_{uH}$	&	$\mathcal{M}(Q_{a\alpha}u_{\dot{a}}H^2_{\beta\gamma}H^{\dagger}_{\dot{\alpha}}) = [12]T^+_{\alpha(\beta\gamma)\dot\alpha}\delta_{a\dot{a}}$ \\
		\hline
	\end{tabular}
	\end{center}
where the group structure $T^+_{\alpha(\beta\gamma)\dot\alpha}\equiv \epsilon_{\alpha \beta} \delta_{\gamma \dot\alpha} +\epsilon_{\alpha \gamma} \delta_{\beta  \dot\alpha} $ symmetric for indices $\beta$ and $\gamma$. 
If we flip the helicity of fermions, we get another three independent amplitude basis in the class of $\mathcal{M}(\psi^- \psi^- \phi^3)$, which is the conjugation of above operator basis. The expressions of these new amplitude basis are obtained by replacing the helicity factor $[12]$ with $\langle 12 \rangle$ in above expressions.

	\item Class $\mathcal{M}(\psi^+\psi^-\phi^2)$ ($\mathcal{O}\sim \psi^2\varphi^2D$):
	
	Note that momentum conservation implies that $ [1|p_3\ket{2} = \frac{1}{2}[1|p_3-p_4\ket{2} $ for $n=4$, and is antisymmetric for the two scalars ($[1|p_3 +p_4| \ket{2}=0$).
Hence there is only one independent term.  	
	\begin{center}
	\begin{tabular}{|l|l|}
		\hline
		Warsaw 			&	Amplitude Basis 	\\
		\hline\hline
		$\mathcal{O}_{He}$	&	$\mathcal{M}(ee^{\dagger}H_{\alpha}H^{\dagger}_{\dot\alpha}) = [1|p_3\ket{2} \delta_{\alpha\dot\alpha}$ \\
		\hline
		$\mathcal{O}_{Hu}$	&	$\mathcal{M}(u_{\dot{a}}u^{\dagger}_{a}H_{\alpha}H^{\dagger}_{\dot\alpha}) =  [1|p_3\ket{2} \delta_{\alpha\dot\alpha}\delta_{a\dot{a}}$ \\
		\hline
		$\mathcal{O}_{Hd}$	&	$\mathcal{M}(d_{\dot{a}}d^{\dagger}_{a}H_{\alpha}H^{\dagger}_{\dot\alpha}) =  [1|p_3\ket{2} \delta_{\alpha\dot\alpha}\delta_{a\dot{a}}$ \\
		\hline
		$\mathcal{O}_{Hud}$	&	$\mathcal{M}(d_{\dot{a}}u^{\dagger}_{a}H^2_{\alpha\beta}) = \frac{1}{2}[1|p_3-p_4\ket{2}\epsilon_{\alpha\beta}\delta_{a\dot{a}}$ \\
$ \mathcal{O}_{Hud}^\dagger$	&	$\mathcal{M}( u_{\dot{a}}d^{\dagger}_{a}H^{\dagger 2}_{\dot \alpha \dot \beta}) = \frac{1}{2}[1|p_3-p_4\ket{2}\epsilon_{\dot\alpha \dot \beta}\delta_{a\dot{a}}$ \\
		\hline
		$\mathcal{O}_{HL}^{(3)} + \frac{3}{4}\mathcal{O}_{HL}^{(1)}$	&	$\mathcal{M}^+(L_{\alpha}L^{\dagger}_{\dot\alpha}H_{\beta}H^{\dagger}_{\dot\beta}) = [1|p_3\ket{2} T^+_{\alpha\beta\dot\alpha\dot\beta}$ \\
		\hline
		$\mathcal{O}_{HL}^{(3)} - \frac{1}{4}\mathcal{O}_{HL}^{(1)}$	&	$\mathcal{M}^-(L_{\alpha}L^{\dagger}_{\dot\alpha}H_{\beta}H^{\dagger}_{\dot\beta}) = [1|p_3\ket{2}T^-_{\alpha\beta\dot\alpha\dot\beta}$ \\
		\hline
		$\mathcal{O}_{HQ}^{(3)} + \frac{3}{4}\mathcal{O}_{HQ}^{(1)}$	&	$\mathcal{M}^+(Q_{a\alpha}Q^{\dagger}_{\dot{a}\dot\alpha}H_{\beta}H^{\dagger}_{\dot\beta}) = [1|p_3\ket{2}T^+_{\alpha\beta\dot\alpha\dot\beta}\delta_{a\dot{a}}$ \\
		\hline
		$\mathcal{O}_{HQ}^{(3)} - \frac{1}{4}\mathcal{O}_{HQ}^{(1)}$	&	$\mathcal{M}^-(Q_{a\alpha}Q^{\dagger}_{\dot{a}\dot\alpha}H_{\beta}H^{\dagger}_{\dot\beta}) = [1|p_3\ket{2} T^-_{\alpha\beta\dot\alpha\dot\beta}\delta_{a\dot{a}}$ \\
		\hline
	\end{tabular}
	\end{center}

	\item Class $\mathcal{M}(A^+\psi^+\psi^+\phi)$ ($\mathcal{O}\sim \psi^2X\varphi$) +h.c.:
	
	\begin{center}
		\begin{tabular}{|l|l|}
			\hline
			Warsaw 			&	Amplitude Basis 	\\
			\hline\hline
			$\mathcal{O}_{eB}$	&	$\mathcal{M}(B^+eL_{\alpha}H^{\dagger}_{\dot\alpha}) = [12][13]\delta_{\alpha\dot\alpha}$ \\
			\hline
			$\mathcal{O}_{dB}$	&	$\mathcal{M}(B^+d_{\dot{a}}Q_{a\alpha}H^{\dagger}_{\dot\alpha}) = [12][13] \delta_{\alpha\dot\alpha}\delta_{a\dot{a}}$ \\
			\hline
			$\mathcal{O}_{dG}$	&	$\mathcal{M}(G^{A+}d_{\dot{b}}Q_{a\alpha}H^{\dagger}_{\dot\alpha}) = [12][13]\delta_{\alpha\dot\alpha}\lambda^A_{a\dot{b}}$ \\
			\hline
			$\mathcal{O}_{eW}$	&	$\mathcal{M}(W^{i+}eL_{\alpha}H^{\dagger}_{\dot\beta}) = [12][13]\tau^i_{\alpha\dot\beta}$ \\
			\hline
			$\mathcal{O}_{dW}$	&	$\mathcal{M}(W^{i+}d_{\dot{a}}Q_{a\alpha}H^{\dagger}_{\dot\beta}) = [12][13]\tau^i_{\alpha\dot\beta}\delta_{a\dot{a}}$ \\
			\hline
			$\mathcal{O}_{uB}$	&	$\mathcal{M}(B^+u_{\dot{a}}Q_{a\alpha}H_{\beta}) = [12][13]\epsilon_{\alpha\beta}\delta_{a\dot{a}}$ \\
			\hline
			$\mathcal{O}_{uW}$	&	$\mathcal{M}(W^{i+}u_{\dot{a}}Q_{a\alpha}H_{\beta}) = [12][13]\tau^{i \beta}_{\alpha}\delta_{a\dot{a}}$ \\
			\hline
			$\mathcal{O}_{uG}$	&	$\mathcal{M}(G^{A+}u_{\dot{b}}Q_{a\alpha}H_{\beta}) = [12][13]\epsilon_{\alpha\beta}\lambda^A_{a\dot{b}}$ \\
			\hline
		\end{tabular}
	\end{center}
	where $\lambda^A_{a\dot{b}}$ is the generator matrix of $SU(3)_c$ and $\tau^{i \beta}_{\alpha} \equiv \tau^{i }_{\alpha \gamma} \epsilon^{\gamma \beta}$. We can get the independent amplitude basis in the class of $\mathcal{M}(A^-\psi^-\psi^-\phi)$ whose expressions can also be obtained by replacing square product $[12][13]$ with angle product $\langle12 \rangle \langle 34 \rangle$.
	\item Class $\mathcal{M}(\psi^+\psi^+\psi^-\psi^-)$ ($\mathcal{O}\sim \bar{L}L\bar{L}L$, $\bar{R}R\bar{R}R$, $\bar{L}L\bar{R}R$, $\bar{L}R\bar{R}L$, $LLRR$) :

	The operators with all left and right handed fermions, and with group factor containing $\epsilon_{abc}$, violate Baryon number conservation.

	\begin{center}
	\begin{tabular}{|l|l|}
		\hline
		Warsaw 			&	Amplitude Basis 	\\
		\hline\hline
		$\mathcal{O}_{qq}^{(3)} + \frac{3}{4}\mathcal{O}_{qq}^{(1)}$	&	$\mathcal{M}^+(Q_{a\alpha}Q_{b\beta}Q^{\dagger}_{\dot{a}\dot\alpha}Q^{\dagger}_{\dot{b}\dot\beta}) = [12]\vev{34}T^+_{\alpha\beta\dot\alpha\dot\beta}T^+_{ab\dot{a}\dot{b}}$ \\
		\hline
		$\mathcal{O}_{qq}^{(3)} - \frac{1}{4}\mathcal{O}_{qq}^{(1)}$	&	$\mathcal{M}^{-}(Q_{a\alpha}Q_{b\beta}Q^{\dagger}_{\dot{a}\dot\alpha}Q^{\dagger}_{\dot{b}\dot\beta}) = [12]\vev{34} T^-_{\alpha\beta\dot\alpha\dot\beta}T^-_{ab\dot{a}\dot{b}}$ \\
		\hline
		$\mathcal{O}_{lq}^{(3)} + \frac{3}{4}\mathcal{O}_{lq}^{(1)}$	&	$\mathcal{M}^{\pm}(Q_{a\alpha}L_{\beta}Q^{\dagger}_{\dot{a}\dot\alpha}L^{\dagger}_{\dot\beta}) = [12]\vev{34}T^+_{\alpha\beta\dot\alpha\dot\beta}\delta_{a\dot{a}}$ \\
		\hline
		$\mathcal{O}_{lq}^{(3)} - \frac{1}{4}\mathcal{O}_{lq}^{(1)}$	&	$\mathcal{M}^{\pm}(Q_{a\alpha}L_{\beta}Q^{\dagger}_{\dot{a}\dot\alpha}L^{\dagger}_{\dot\beta}) = [12]\vev{34}T^-_{\alpha\beta\dot\alpha\dot\beta}\delta_{a\dot{a}}$ \\
		\hline
		$\mathcal{O}_{ll}$	&	$\mathcal{M}(L_{\alpha}L_{\beta}L^{\dagger}_{\dot\alpha}L^{\dagger}_{\dot\beta}) = [12]\vev{34} T^+_{\alpha\beta\dot\alpha\dot\beta}$ \\
		\hline
		$\mathcal{O}_{qu}^{(8)} + \frac{2}{3}\mathcal{O}_{qu}^{(1)}$	&	$\mathcal{M}^{\pm}(Q_{a\alpha}u_{\dot{b}}Q^{\dagger}_{\dot{a}\dot\alpha}u^{\dagger}_{b}) = [12]\vev{34}\delta_{\alpha\dot\alpha}T^+_{ab\dot{a}\dot{b}}$ \\
		\hline
		$\mathcal{O}_{qu}^{(8)} - \frac{1}{3}\mathcal{O}_{qu}^{(1)}$	&	$\mathcal{M}^{\pm}(Q_{a\alpha}u_{\dot{b}}Q^{\dagger}_{\dot{a}\dot\alpha}u^{\dagger}_{b}) = [12]\vev{34}\delta_{\alpha\dot\alpha}T^-_{ab\dot{a}\dot{b}}$ \\
		\hline
		$\mathcal{O}_{qd}^{(8)} + \frac{2}{3}\mathcal{O}_{qd}^{(1)}$	&	$\mathcal{M}^{\pm}(Q_{a\alpha}d_{\dot{b}}Q^{\dagger}_{\dot{a}\dot\alpha}d^{\dagger}_{b}) = [12]\vev{34}\delta_{\alpha\dot\alpha}T^+_{ab\dot{a}\dot{b}}$ \\
		\hline
		$\mathcal{O}_{qd}^{(8)} - \frac{1}{3}\mathcal{O}_{qd}^{(1)}$	&	$\mathcal{M}^{\pm}(Q_{a\alpha}d_{\dot{b}}Q^{\dagger}_{\dot{a}\dot\alpha}d^{\dagger}_{b}) = [12]\vev{34}\delta_{\alpha\dot\alpha}T^-_{ab\dot{a}\dot{b}}$ \\
		\hline
		$\mathcal{O}_{ud}^{(8)} + \frac{2}{3}\mathcal{O}_{ud}^{(1)}$	&	$\mathcal{M}^{\pm}(u_{\dot{a}}d_{\dot{b}}u^{\dagger}_{a}d^{\dagger}_{b}) = [12]\vev{34}T^+_{ab\dot{a}\dot{b}}$ \\
		\hline
		$\mathcal{O}_{ud}^{(8)} - \frac{1}{3}\mathcal{O}_{ud}^{(1)}$	&	$\mathcal{M}^{\pm}(u_{\dot{a}}d_{\dot{b}}u^{\dagger}_{a}d^{\dagger}_{b}) = [12]\vev{34}T^-_{ab\dot{a}\dot{b}}$ \\
		\hline
		$\mathcal{O}_{uu}$	&	$\mathcal{M}(u_{\dot{a}}u_{\dot{b}}u^{\dagger}_{a}u^{\dagger}_{b}) = [12]\vev{34}T^+_{ab\dot{a}\dot{b}}$ \\
		\hline
		$\mathcal{O}_{dd}$	&	$\mathcal{M}(d_{\dot{a}}d_{\dot{b}}d^{\dagger}_{a}d^{\dagger}_{b}) = [12]\vev{34}T^+_{ab\dot{a}\dot{b}}$ \\
		\hline
		$\mathcal{O}_{lu}$	&	$\mathcal{M}(L_{\alpha}u_{\dot{a}}L^{\dagger}_{\dot\alpha}u^{\dagger}_a) = [12]\vev{34}\delta_{\alpha\dot\alpha}\delta_{a\dot{a}}$ \\
		\hline
		$\mathcal{O}_{ld}$	&	$\mathcal{M}(L_{\alpha}d_{\dot{a}}L^{\dagger}_{\dot\alpha}d^{\dagger}_a) = [12]\vev{34}\delta_{\alpha\dot\alpha}\delta_{a\dot{a}}$ \\
		\hline
		$\mathcal{O}_{qe}$	&	$\mathcal{M}(Q_{a\alpha}e Q^{\dagger}_{\dot{a}\dot\alpha}e^{\dagger}) = [12]\vev{34}\delta_{\alpha\dot\alpha}\delta_{a\dot{a}}$ \\
		\hline
		$\mathcal{O}_{ledq}$	&	$\mathcal{M}(Q_{a\alpha}d_{\dot{a}}L^{\dagger}_{\dot\alpha}e^{\dagger}) = [12]\vev{34}\delta_{\alpha\dot\alpha}\delta_{a\dot{a}}$ \\
     $\mathcal{O}_{ledq}^\dagger$ &	$\mathcal{M}(L_{\alpha}e Q_{\dot a \dot \alpha}^\dagger d_{a}^\dagger ) = [12]\vev{34}\delta_{\alpha\dot\alpha}\delta_{a\dot{a}}$ \\
		\hline
		$\mathcal{O}_{le}$	&	$\mathcal{M}(L_{\alpha}e L^{\dagger}_{\dot\alpha}e^{\dagger}) =[12]\vev{34}\delta_{\alpha\dot\alpha}$ \\
		\hline
		$\mathcal{O}_{eu}$	&	$\mathcal{M}(eu_{\dot{a}}e^{\dagger}u^{\dagger}_a) = [12]\vev{34}\delta_{a\dot{a}}$ \\
		\hline
		$\mathcal{O}_{ed}$	&	$\mathcal{M}(ed_{\dot{a}}e^{\dagger}d^{\dagger}_a) =[12]\vev{34}\delta_{a\dot{a}}$ \\
		\hline
		$\mathcal{O}_{ee}$	&	$\mathcal{M}(e^2e^{\dagger2}) = [12]\vev{34}$ \\
		\hline
		$\mathcal{O}_{duq}$	&	$\mathcal{M}(Q_{a\alpha}L_{\beta}u^{\dagger}_{b}d^{\dagger}_{c}) =[12]\vev{34}\epsilon_{\alpha\beta}\epsilon_{abc}$ \\
        $\mathcal{O}_{duq}^\dagger$	&	$\mathcal{M}(u_{\dot b}d_{\dot c} Q_{\dot a \dot \alpha}^\dagger L_{\dot \beta}^\dagger) =[12]\vev{34}\epsilon_{\dot \alpha \dot \beta}\epsilon_{\dot a \dot b \dot c}$ \\

		\hline
		$\mathcal{O}_{qqu}$	&	$\mathcal{M}(Q_{a\alpha}Q_{b\beta}u^{\dagger}_{c}e^{\dagger}) = [12]\vev{34}\epsilon_{\alpha\beta}\epsilon_{abc}$ \\
   $ \mathcal{O}_{qqu}^\dagger$	&	$\mathcal{M}(u_{\dot c}e Q_{\dot a \dot \alpha}^\dagger Q_{\dot b \dot\beta}^\dagger) = [12]\vev{34}\epsilon_{\dot \alpha \dot \beta}\epsilon_{\dot a \dot b \dot c}$ \\
		\hline
	\end{tabular}
\end{center}

\item[8] Class $\mathcal{M}(\psi^+\psi^+\psi^+\psi^+)$ ($\mathcal{O}\sim \bar{L}R\bar{L}R$, $LLLL$, $RRRR$) + h.c.:

Notice that $f(\psi^+\psi^+\psi^+\psi^+)$ has two choices. We define combinations $f^\pm \equiv [13][24] \pm [23][14]$ with specific permutation symmetries. $\mathcal{O}_{lequ}^{(3)}$ is not the weak current interaction, but defined as different spinor contractions $\sigma^{\mu\nu}\sigma_{\mu\nu}$. The failure of a unified notation in Warsaw basis proves the advantage of the amplitude basis as a systematic classification.
	
\begin{center}
	\begin{tabular}{|l|l|}
		\hline
		Warsaw 			&	Amplitude Basis 	\\
		\hline\hline
		$\mathcal{O}_{quqd}^{(8)} + \frac{2}{3}\mathcal{O}_{quqd}^{(1)}$	&	$\mathcal{M}^+(Q_{a\alpha}Q_{b\beta}u_{\dot{a}}d_{\dot{b}}) = f^-\epsilon_{\alpha\beta}T^-_{ab\dot{a}\dot{b}}$ \\
		\hline
		$\mathcal{O}_{quqd}^{(8)} - \frac{1}{3}\mathcal{O}_{quqd}^{(1)}$	&	$\mathcal{M}^-(Q_{a\alpha}Q_{b\beta}u_{\dot{a}}d_{\dot{b}}) = f^+ \epsilon_{\alpha\beta}T^+_{ab\dot{a}\dot{b}}$ \\
		\hline
		$-\frac{1}{4}\mathcal{O}_{lequ}^{(3)} + \mathcal{O}_{lequ}^{(1)}$	&	$\mathcal{M}^+(L_{\alpha}Q_{a\beta}u_{\dot{a}}e) = f^+\epsilon_{\alpha\beta}\delta_{a\dot{a}}$ \\
		\hline
		$-\frac{1}{4}\mathcal{O}_{lequ}^{(3)} - 3\mathcal{O}_{lequ}^{(1)}$	&	$\mathcal{M}^-(L_{\alpha}Q_{a\beta}u_{\dot{a}}e) = f^-\epsilon_{\alpha\beta}\delta_{a\dot{a}}$ \\
		\hline
		$\mathcal{O}_{qqq}$	&	$\mathcal{M}(Q_{a\alpha}Q_{b\beta}Q_{c\gamma}L_{\delta}) = f^-T^-_{\alpha\beta\gamma\delta}\epsilon_{abc}$ \\
		\hline
		$\mathcal{O}_{duu}$	&	$\mathcal{M}(u^2_{\dot{a}\dot{b}}d_{\dot{c}}e) = f^+\epsilon_{\dot{a}\dot{b}\dot{c}}$ \\
		\hline
	\end{tabular}
\end{center}
Following the same procedure, we can obtain the amplitude basis in the class of $\mathcal{M}(\psi^-\psi^-\psi^-\psi^-)$ by replacing square product with angle product.
	
\end{enumerate}

Summing up all 8 classes of amplitude basis, we get $3+8+4+6+9+16+12+26=84$ basis (hermitian conjugates are counted separately), recovering the well known result.

\section{Conclusion and Outlook}

In this letter, we propose a novel way of writing down all independent effective operators from the unfactorizable on-shell amplitudes. This particular basis is referred as the amplitude basis since all operators are in one to one correspondence with the on-shell amplitudes. We provide the general rules to construct those primary amplitudes and classify them by the external legs and helicity assignments so that all operators in the amplitude basis can be enumerated systematically for a given dimension. Then we further demonstrate how to use our method to generate all independent dim-$5$ and dim-$6$ operators in SMEFT while respecting the SM gauge symmetry and spin-statistics constrains. Interestingly, we find that operators in our amplitude basis for $d=6$ SMEFT is the well known Warsaw basis, except for some linear combinations. Our method starts from the on-shell amplitudes thus it is naturally convenient for EFT calculation and free from redundancies connected by EOM and IBP.

Our result here is only a small tip of the iceberg for the on shell effective field theory. There are various interesting aspects that can be done or will be finished very soon (some related applications are discussed in Ref~\cite{Cheung:2015aba,Azatov:2016sqh}). The procedure can be applied to more sophisticated cases together with tools to deal with tensor structures in $d=7$, $8$ SMEFT~\cite{MSXZ2019}. Applications to specific  processes can be demonstrated as Ref.~\cite{Shadmi:2018xan}. The SMEFT is a massless case, applications to the EFT with massive particles~\cite{Arkani-Hamed:2017jhn} are under investigation. The current setup can be encoded into the computation of Wilson coefficients of the amplitude basis if we know the underlying theory. Applications to other types of EFT and related concepts may also spark off intriguing results.

\section*{Acknowledgements}

We thank Song He and Hui Luo for useful discussions and comments. J.S. thanks the hospitality of HKUST Jockey Club Institute for Advanced Study while working on this project. T.M. is supported in part by project Y6Y2581B11 supported by 2016 National Postdoctoral Program for Innovative Talents. J.S. is supported by the National Natural Science Foundation of China (NSFC) under grant No.11847612, No.11690022, No.11851302, No.11675243 and No.11761141011 and also supported by the Strategic Priority Research Program of the Chinese Academy of Sciences under grant No.XDB21010200 and No.XDB23000000. M.L.X. is supported by 2019 the International Postdoctoral Exchange Fellowship Program.

\appendix

\section{Notation}
In this section we list the conventions throughout this work.

\subsection{Conventions for spinor helicity formalism}
\label{sub:spinor}
Since the Lorentz group $SO(3,1)$ is isomorphism with $SU(2)_L\times  SU(2)_R$, the four-vector momentum $p_\mu$ can be mapped into a two-by-two matrix via
\bea
p_{\alpha \dot \alpha} = p_\mu \sigma^\mu_{\alpha \dot \alpha}
\eea
where $\sigma_\mu = (1,\sigma^i)$ is a four-vector of Pauli matrices and the undotted and dotted indices transform under the usual spinor representations of the Lorentz group. We can find the determinant of $p_{\alpha \dot \alpha}$ is Lorentz scalar
\bea
\text{det}[p]=p_\mu p^\mu,
\eea
which vanishes for massless on-shell particle. So the vanishing determinant of massless on-shell particle momentum indicates that $p_{\alpha \dot \alpha}$ is a two-by-two matrix of at most rank one, which can be written as the outer product of two two-component objects which are called spinors
\bea \label{eq:decomposition}
p_{\alpha \dot \alpha} =-|p]_{\alpha} \langle p| _{\dot\alpha}.
\eea
Given two massless particles $i$ and $j$, we can define Lorentz invariant building blocks of spinor helicity formalism
\bea
\langle ij \rangle =\epsilon^{\alpha \beta} |i\rangle_\alpha |j\rangle_\beta   \quad [ ij ] =\epsilon^{\dot \alpha  \dot \beta} |i]_{\dot \alpha} |j]_{\dot \beta}
\eea
where we use the short-hand notation $|i\rangle \equiv |p_i\rangle$ ($|i] \equiv |p_i ]$) and $\epsilon^{\alpha \beta}$ is $2$-index Levi-Civita symbols. The  Mandelstam invariants can be written in terms of these objects
\bea
s_{ij}=(p_i +p_j)^2 =2p_i.p_j =\langle ij \rangle [ ij ].
\eea

Because spinors are two dimension objects, one can always write a spinor as a linear combination of two linearly independent spinors and thus have the identity
\bea
[ij][kl] +[ik][lj]+[il][jk]=0,
\eea
which is known as the Schouten identity.

 The lightlike momentum decomposition in Eq.~\ref{eq:decomposition} is invariant under the scaling
 \bea \label{eq:little_group}
| p \rangle \to t | p \rangle \quad [ p|  \to t^{-1} [p|.
 \eea
For real momentum, the scaling factor $t$ is just a pure phase. So the transformation in Eq.~(\ref{eq:little_group}) corresponds to the $SO(2)$ little group transformation of the lightlike momentum, which is called little group scaling.

For an on-shell amplitude, the $i$th external leg with helicity $h_i$ scales as $t^{-2h_i}$ and neither propagators or vertices can scale under little group. So the on-shell amplitude transform homogeneously under little group scaling
\bea
A_n(1^{h_1},2^{h_2},..., n^{h_n} ) \to \prod_{i} t^{2h_i}A_n(1^{h_1},2^{h_2},..., n^{h_n} ).
\eea
The transformation of the on-shell scattering amplitudes under little group scaling can help determine the little group weight function $f$~(see the review~\cite{Cheung:2017pzi, Elvang:2013cua}).

\subsection{Conventions for SM fields and gauge symmetry}
\label{sub:SM_notation}
In this section, we list the notations of SM fields and their gauge symmetry indices in Tab.~(\ref{tab:SM}), where all fermions are listed as left-handed Weyl fermions.
\begin{table}
	\centering
	\begin{tabular}{c|ccc}
							&	$SU(3)_c$			&	$SU(2)_L$	&	$U(1)_Y$ \\
		\hline
		$G_A^\pm$				&	$\mathbf{3}$		&	$\mathbf{1}$&	0		\\
		$W_i^\pm$				&	$\mathbf{1}$		&	$\mathbf{3}$&	0		\\
		$B^\pm$				&	$\mathbf{1}$		&	$\mathbf{1}$&	0		\\
		$Q_{a\alpha}$		&	$\mathbf{3}$		&	$\mathbf{2}$&	1/6		\\
		$u_{\dot{a}}$		&	$\mathbf{\bar{3}}$	&	$\mathbf{1}$&	-2/3	\\
		$d_{\dot{a}}$		&	$\mathbf{\bar{3}}$	&	$\mathbf{1}$&	1/3		\\
		$L_{\alpha}$		&	$\mathbf{1}$		&	$\mathbf{2}$&	-1/2	\\
		$e$					&	$\mathbf{1}$		&	$\mathbf{1}$&	1		\\
		$H_{\alpha}$		&	$\mathbf{1}$		&	$\mathbf{2}$&	1/2		\\
		\hline
	\end{tabular}
	\caption{Standard Model particle content is listed according to their representations under gauge group $SU(3)_c \times SU(2)_L \times U(1)_Y$. All fermions are in the form of left hand.}\label{tab:SM}
\end{table}
We require that the anti-fundamental representation of $SU(3)_c$ are denoted by dotted letters $\dot a, \dot b,...$ and the indices of the conjugate of $SU(2)_L$ doublets of SM left-handed fermions and Higgs doublet with hypercharge $1/2$ are denoted by dotted Greece letter $\dot \alpha, \dot \beta, ...$.

\section{Warsaw Basis}
We list all the standard Warsaw basis operators below.
We mostly keep the notations used in this paper, and for consistency we label the right handed fermions as $u_R = Cu^*$, $d_R = Cd^*$, $e_R = Ce^*$, where $C = i\sigma^2$ for Weyl spinors and $C = i\gamma^0\gamma^2$ for Dirac spinors. We use four-component Dirac spinors here as the original Warsaw paper did. $\sigma^{\mu\nu} = [\gamma^{\mu},\gamma^{\nu}]$ and $T^A = \lambda^A/2$.

\begin{center}
	\begin{tabular}{|c|c|c|c|}
		\hline
		\multicolumn{2}{|c}{$X^3$} &	\multicolumn{2}{|c|}{$X^2\varphi^2$} \\
		\hline
		$\mathcal{O}_G$	&	$f^{ABC}G^A_{\mu\nu}G^B_{\nu\rho}G^C_{\rho\mu}$	&	
		$\mathcal{O}_{H\stackrel{(\sim)}{B}}$	&	$H^{\dagger}H\stackrel{(\sim)}{B}_{\mu\nu}B^{\mu\nu}$	\\
		$\mathcal{O}_{\tilde{G}}$	&	$f^{ABC}\tilde{G}^A_{\mu\nu}G^B_{\nu\rho}G^C_{\rho\mu}$	&	
		$\mathcal{O}_{H\stackrel{(\sim)}{W}B}$	&	$H^{\dagger}\tau^iH\stackrel{(\sim)}{W^i}_{\mu\nu}B^{\mu\nu}$	\\
		$\mathcal{O}_W$	&	$\epsilon^{ijk}W^i_{\mu\nu}W^j_{\nu\rho}W^k_{\rho\mu}$	&	
		$\mathcal{O}_{H\stackrel{(\sim)}{W}}$	&	$H^{\dagger}H\stackrel{(\sim)}{W^i}_{\mu\nu}W^{i\mu\nu}$	\\
		$\mathcal{O}_{\tilde{W}}$	&	$\epsilon^{ijk}\tilde{W}^i_{\mu\nu}W^j_{\nu\rho}W^k_{\rho\mu}$	&	
		$\mathcal{O}_{H\stackrel{(\sim)}{G}}$	&	$H^{\dagger}H\stackrel{(\sim)}{G^A}_{\mu\nu}G^{A\mu\nu}$	\\
		\hline
		\multicolumn{2}{|c}{$\varphi^6$ and $\varphi^4D^2$}	&	\multicolumn{2}{|c|}{$\psi^2\varphi^3$}\\
		\hline
		$\mathcal{O}_H$	&	$(H^{\dagger}H)^3$	&	
		$\mathcal{O}_{eH}$	&	$(H^{\dagger}H)(\bar{L} e_R H)$	\\
		$\mathcal{O}_{H\Dbox}$	&	$(H^{\dagger}H)\Dbox(H^{\dagger}H)$	&	
		$\mathcal{O}_{uH}$	&	$(H^{\dagger}H)(\bar{Q} u_R \tilde{H})$	\\
		$\mathcal{O}_{HD}$	&	$|H^{\dagger}D_{\mu}H|^2$	&
		$\mathcal{O}_{eH}$	&		$(H^{\dagger}H)(\bar{Q} d_R H)$	\\
		\hline

\multicolumn{2}{|c}{$\psi^2X\varphi$} &	\multicolumn{2}{|c|}{$\psi^2\varphi^2D$} \\
		\hline
		$\mathcal{O}_{eB}$	&	$(\bar{L}\sigma^{\mu\nu}e_R)HB_{\mu\nu}$	&	
		$\mathcal{O}_{He}$	&	$(H^{\dagger}i\overleftrightarrow{D}_{\mu}H)(\bar{e}_R\gamma^{\mu}e_R)$	\\
		$\mathcal{O}_{dB}$	&	$(\bar{Q}\sigma^{\mu\nu}d_R)HB_{\mu\nu}$	&	
		$\mathcal{O}_{Hu}$	&	$(H^{\dagger}i\overleftrightarrow{D}_{\mu}H)(\bar{u}_R\gamma^{\mu}u_R)$	\\
		$\mathcal{O}_{dG}$	&	$(\bar{Q}\frac{\lambda^A}{2}\sigma^{\mu\nu}d_R)HG^A_{\mu\nu}$	&	
		$\mathcal{O}_{Hd}$	&	$(H^{\dagger}i\overleftrightarrow{D}_{\mu}H)(\bar{d}_R\gamma^{\mu}d_R)$	\\
		$\mathcal{O}_{eW}$	&	$(\bar{L}\sigma^{\mu\nu}e_R)\tau^iHW^i_{\mu\nu}$	&	
		$\mathcal{O}_{Hud}$	&	$(\tilde{H}^{\dagger}iD_{\mu}H)(\bar{u}_R\gamma^{\mu}d_R)$	\\
		$\mathcal{O}_{dW}$	&	$(\bar{Q}\sigma^{\mu\nu}d_R)\tau^iHW^i_{\mu\nu}$	&	
		$\mathcal{O}^{(1)}_{Hl}$	&	$(\tilde{H}^{\dagger}iD_{\mu}H)(\bar{L}\gamma^{\mu}L)$	\\
		$\mathcal{O}_{uB}$	&	$(\bar{Q}\sigma^{\mu\nu}u_R)\tilde{H}B_{\mu\nu}$	&	
		$\mathcal{O}^{(3)}_{Hl}$	&	$(\tilde{H}^{\dagger}iD^i_{\mu}H)(\bar{L}\tau^i\gamma^{\mu}L)$	\\
		$\mathcal{O}_{uW}$	&	$(\bar{Q}\sigma^{\mu\nu}u_R)\tau^i\tilde{H}W^i_{\mu\nu}$	&	
		$\mathcal{O}^{(1)}_{Hq}$	&	$(\tilde{H}^{\dagger}iD_{\mu}H)(\bar{Q}\gamma^{\mu}Q)$	\\
		$\mathcal{O}_{uG}$	&	$(\bar{Q}\frac{\lambda^A}{2}\sigma^{\mu\nu}u_R)\tilde{H}G^A_{\mu\nu}$	&	
		$\mathcal{O}^{(3)}_{Hq}$	&	$(\tilde{H}^{\dagger}iD^i_{\mu}H)(\bar{Q}\tau^i\gamma^{\mu}Q)$	\\
		\hline
\multicolumn{2}{|c|}{$(\bar{L}R)(\bar{L}R)$} &	\multicolumn{2}{c|}{$LLLL$ ($B\!\!\!\!/$)} \\
		\hline
		$\mathcal{O}_{quqd}^{(1)}$	&	$(\bar{Q}u_R)\epsilon(\bar{Q}d_R)$	&	
		$\mathcal{O}_{qqq}$	&	$\epsilon_{abc}(Q^a \epsilon CQ^b)(Q^c\epsilon CL)$	\\
		\cline{3-4}
		$\mathcal{O}_{quqd}^{(8)}$	&	$(\bar{Q}T^Au_R)\epsilon(\bar{Q}T^Ad_R)$	&	
		\multicolumn{2}{c|}{$RRRR$ ($B\!\!\!\!/$)}	\\
		\cline{3-4}
		$\mathcal{O}_{lequ}^{(1)}$	&	$(\bar{L}e_R)\epsilon(\bar{Q}d_R)$	&	
		$\mathcal{O}_{duu}$	&	$\epsilon_{abc}(d_R^aCu_R^b)(u_R^cCe_R)$	\\
		$\mathcal{O}_{lequ}^{(3)}$	&	$(\bar{L}\sigma_{\mu\nu}e_R)\epsilon(\bar{Q}\sigma^{\mu\nu}u_R)$	&	
		&	\\
		\hline
	\end{tabular}
\end{center}

\begin{center}
	\begin{tabular}{|c|c|c|c|}
		\hline
		\multicolumn{2}{|c}{$(\bar{L}L)(\bar{L}L)$} &	\multicolumn{2}{|c|}{$(\bar{L}L)(\bar{R}R)$} \\
		\hline
		$\mathcal{O}_{ll}$	&	$(\bar{L}\gamma_{\mu}L)(\bar{L}\gamma^{\mu}L)$	&	
		$\mathcal{O}_{le}$	&	$(\bar{L}\gamma_{\mu}L)(\bar{e}_R\gamma^{\mu}e_R)$	\\
		$\mathcal{O}_{qq}^{(1)}$	&	$(\bar{Q}\gamma_{\mu}Q)(\bar{Q}\gamma^{\mu}Q)$	&	
		$\mathcal{O}_{lu}$	&	$(\bar{L}\gamma_{\mu}L)(\bar{u}_R\gamma^{\mu}u_R)$	\\
		$\mathcal{O}_{qq}^{(3)}$	&	$(\bar{Q}\gamma_{\mu}\tau^iQ)(\bar{Q}\gamma^{\mu}\tau^iQ)$	&	
		$\mathcal{O}_{ld}$	&	$(\bar{L}\gamma_{\mu}L)(\bar{d}_R\gamma^{\mu}d_R)$	\\
		$\mathcal{O}_{lq}^{(1)}$	&	$(\bar{L}\gamma_{\mu}L)(\bar{Q}\gamma^{\mu}Q)$	&	
		$\mathcal{O}_{qe}$	&	$(\bar{Q}\gamma_{\mu}Q)(\bar{e}_R\gamma^{\mu}e_R)$	\\
		$\mathcal{O}_{lq}^{(3)}$	&	$(\bar{L}\gamma_{\mu}\tau^iL)(\bar{Q}\gamma^{\mu}\tau^iQ)$	&	
		$\mathcal{O}_{qu}^{(1)}$	&	$(\bar{Q}\gamma_{\mu}Q)(\bar{u}_R\gamma^{\mu}u_R)$	\\
		\cline{1-2}
		\multicolumn{2}{|c|}{$(\bar{R}R)(\bar{R}R)$} &	
		$\mathcal{O}_{qu}^{(8)}$	&	$(\bar{Q}\gamma_{\mu}\frac{\lambda^A}{2}Q)(\bar{u}_R\gamma^{\mu}\frac{\lambda^A}{2}u_R)$	\\
		\cline{1-2}
		$\mathcal{O}_{ee}$	&	$(\bar{e}\gamma_{\mu}e)(\bar{e}\gamma^{\mu}e)$	&	
		$\mathcal{O}_{qd}^{(1)}$	&	$(\bar{Q}\gamma_{\mu}Q)(\bar{d}_R\gamma^{\mu}d_R)$	\\
		$\mathcal{O}_{uu}$	&	$(\bar{u}\gamma_{\mu}u)(\bar{u}\gamma^{\mu}u)$	&	
		$\mathcal{O}_{qd}^{(8)}$	&	$(\bar{Q}\gamma_{\mu}\frac{\lambda^A}{2}Q)  (\bar{d}_R\gamma^{\mu}\frac{\lambda^A}{2}d_R)$	\\
		\cline{3-4}
		$\mathcal{O}_{dd}$	&	$(\bar{d}_R\gamma_{\mu}d_R)(\bar{d}_R\gamma^{\mu}d_R)$	&	
		\multicolumn{2}{|c|}{$(\bar{L}R)(\bar{R}L)$}	\\
		\cline{3-4}
		$\mathcal{O}_{eu}$	&	$(\bar{e}_R\gamma_{\mu}e_R)(\bar{u}_R\gamma^{\mu}u_R)$	&	
		$\mathcal{O}_{ledq}$	&	$(\bar{L}e_R)(\bar{d}_RQ)$	\\
		\cline{3-4}
		$\mathcal{O}_{ed}$	&	$(\bar{e}_R\gamma_{\mu}e_R)(\bar{d}_R\gamma^{\mu}d_R)$	&		
		\multicolumn{2}{|c|}{$LLRR$ ($B\!\!\!\!/$)}	\\
		\cline{3-4}
		$\mathcal{O}^{(1)}_{ud}$	&	$(\bar{u}_R\gamma_{\mu}u_R)(\bar{d}_R\gamma^{\mu}d_R)$	&	
		$\mathcal{O}_{duq}$	&	$\epsilon_{abc}(d_R^aCu_R^b)(Q^c\epsilon CL)$	\\
		$\mathcal{O}^{(8)}_{ud}$	&	$(\bar{u}_R\gamma_{\mu}\frac{\lambda^A}{2}u_R)$	&	$\mathcal{O}_{qqu}$	&	$\epsilon_{abc}(Q^a\epsilon CQ^b)(u_R^cCe_R)$	\\
			&	$(\bar{d}_R\gamma^{\mu}\frac{\lambda^A}{2}d_R)$	&	&	\\
		
		\hline
	\end{tabular}
\end{center}

\section{Example}
\label{ap:ex}

\tikzset{
	photon/.style={decorate, decoration={snake}, draw=black,solid },
	scalar/.style={draw=black,dashed },
	fermion/.style={draw=black, postaction={decorate},decoration={markings,mark=at position .55 with {\arrow{<}}}},
	point/.style={
		thick,
		draw=gray,
		cross out,
		inner sep=0pt,
		minimum width=4pt,
		minimum height=4pt,
	},
}
As an example, let's look at the elastic scattering $W^{a,+}\pi^b \to W^{a,-}\pi^b$ where $\pi$'s are Goldstones in the Higgs doublet, and $a$, $b$ are the group indices. ``$+$" indicates positive helicity and we follow the convention that all external momentums are going outward so the incoming $W^+$ is turned into an outgoing $W^-$ in the elastic scattering. We are interested in how the dim-6 operators contribute to this process. If we use the SILH basis~\cite{Giudice:2007fh}, naively, the following two operators would contribute:
\eq{\label{eq:silh}
	& \mathcal{O}_{W} = \frac{ig}{2}(H^{\dagger}\tau^i\overleftrightarrow{D}_{\mu}H)(D_{\nu}W^{\mu\nu})^i ,\\
	& \mathcal{O}_{HW} = ig(D_{\mu}H)^{\dagger}\tau^i(D_{\nu}H)W^{i\mu\nu}.
}
They contribute via two Feynman diagrams as shown in Fig.~\ref{fig:feyn}, but explicit computation shows that they cancel each other. Hence the final answer is that there is no dim-6 contribution to this elastic scattering.

\begin{figure}[t]
\centering
\begin{tikzpicture}
	\draw[photon] (-1,1.2) -- (0,0) node[left]{\(\mathcal{O}\)};
	\draw[photon] (1,1.2) -- (0,0);
	\draw[scalar] (-1,-1.2) -- (0,-0);
	\draw[scalar] (1,-1.2) -- (0,-0);
\end{tikzpicture}
\hspace{0.5cm}
\begin{tikzpicture}
	\draw[photon] (-1,1.2) -- (0,0.6);
	\draw[photon] (1,1.2) -- (0,0.6);
	\draw[photon] (0,0.6) -- (0,-0.6) node[left]{\(\mathcal{O}\)};
	\draw[scalar] (-1,-1.2) -- (0,-0.6);
	\draw[scalar] (1,-1.2) -- (0,-0.6);
\end{tikzpicture}
\caption{Two Feynman diagrams that contribute to $W^{a,+}\pi^b \to W^{a,+}\pi^b$ from effective operators.}
\label{fig:feyn}
\end{figure}
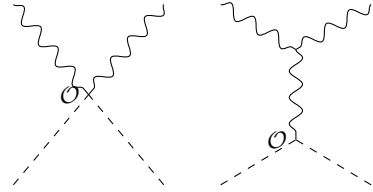

In our amplitude basis language, this conclusion can be made in a much simpler way without any cancelation in Fig~\ref{fig:feyn}. The elastic scattering process is of type $f(F^-F^+\phi^2)$, whose primary amplitude is $[2|p_3\ket{1}^2$ which is dimension $8$. There is also no way to construct on-shell amplitudes recursively from $d=6$, hence no overall on-shell contribution is from dimension 6 operators.


\begin{thebibliography}{99}

\bibitem{Ginzburg:1950sr}
  V.~L.~Ginzburg and L.~D.~Landau,
  Zh.\ Eksp.\ Teor.\ Fiz.\  {\bf 20}, 1064 (1950).


\bibitem{AQHE}
  R.~B.~Laughlin,
  Phys.\ Rev.\ Lett.\  {\bf 50}, 1395 (1983).
  doi:10.1103/PhysRevLett.50.1395

\bibitem{Weinberg:1968de}
  S.~Weinberg,
  Phys.\ Rev.\  {\bf 166}, 1568 (1968).
  doi:10.1103/PhysRev.166.1568

\bibitem{Buchmuller:1985jz}
  W.~Buchmuller and D.~Wyler,
  Nucl.\ Phys.\ B {\bf 268}, 621 (1986).
  doi:10.1016/0550-3213(86)90262-2

\bibitem{Grzadkowski:2010es}
  B.~Grzadkowski, M.~Iskrzynski, M.~Misiak and J.~Rosiek,
  JHEP {\bf 1010}, 085 (2010)
  doi:10.1007/JHEP10(2010)085
  [arXiv:1008.4884 [hep-ph]].


\bibitem{Lehman:2014jma}
  L.~Lehman,
  Phys.\ Rev.\ D {\bf 90}, no. 12, 125023 (2014)
  doi:10.1103/PhysRevD.90.125023
  [arXiv:1410.4193 [hep-ph]].



\bibitem{Liao:2016hru}
  Y.~Liao and X.~D.~Ma,
  JHEP {\bf 1611}, 043 (2016)
  doi:10.1007/JHEP11(2016)043
  [arXiv:1607.07309 [hep-ph]].

\bibitem{Liao:2016qyd}
  Y.~Liao and X.~D.~Ma,
  Phys.\ Rev.\ D {\bf 96}, no. 1, 015012 (2017)
  doi:10.1103/PhysRevD.96.015012
  [arXiv:1612.04527 [hep-ph]].


\bibitem{Jenkins:2009dy}
  E.~E.~Jenkins and A.~V.~Manohar,
  JHEP {\bf 0910}, 094 (2009)
  doi:10.1088/1126-6708/2009/10/094
  [arXiv:0907.4763 [hep-ph]].

\bibitem{Lehman:2015via}
  L.~Lehman and A.~Martin,
  Phys.\ Rev.\ D {\bf 91}, 105014 (2015)
  doi:10.1103/PhysRevD.91.105014
  [arXiv:1503.07537 [hep-ph]].

\bibitem{Lehman:2015coa}
  L.~Lehman and A.~Martin,
  JHEP {\bf 1602}, 081 (2016)
  doi:10.1007/JHEP02(2016)081
  [arXiv:1510.00372 [hep-ph]].

\bibitem{Henning:2015alf}
  B.~Henning, X.~Lu, T.~Melia and H.~Murayama,
  JHEP {\bf 1708}, 016 (2017)
  doi:10.1007/JHEP08(2017)016
  [arXiv:1512.03433 [hep-ph]].

\bibitem{Henning:2017fpj}
  B.~Henning, X.~Lu, T.~Melia and H.~Murayama,
  JHEP {\bf 1710}, 199 (2017)
  doi:10.1007/JHEP10(2017)199
  [arXiv:1706.08520 [hep-th]].


\bibitem{Gripaios:2018zrz}
  B.~Gripaios and D.~Sutherland,
  arXiv:1807.07546 [hep-ph].

\bibitem{Witten:2003nn}
  E.~Witten,
  Commun.\ Math.\ Phys.\  {\bf 252}, 189 (2004)
  doi:10.1007/s00220-004-1187-3
  [hep-th/0312171].

\bibitem{Shadmi:2018xan}
  Y.~Shadmi and Y.~Weiss,
  arXiv:1809.09644 [hep-ph].

\bibitem{Cohen:2010mi}
  T.~Cohen, H.~Elvang and M.~Kiermaier,
  JHEP {\bf 1104}, 053 (2011)
  doi:10.1007/JHEP04(2011)053
  [arXiv:1010.0257 [hep-th]].

\bibitem{Elvang:2010jv}
  H.~Elvang, D.~Z.~Freedman and M.~Kiermaier,
  JHEP {\bf 1011}, 016 (2010)
  doi:10.1007/JHEP11(2010)016
  [arXiv:1003.5018 [hep-th]].
  
  
\bibitem{Cheung:2015aba}
  C.~Cheung and C.~H.~Shen,
  Phys.\ Rev.\ Lett.\  {\bf 115}, no. 7, 071601 (2015)
  doi:10.1103/PhysRevLett.115.071601
  [arXiv:1505.01844 [hep-ph]].
  
  
\bibitem{Azatov:2016sqh}
  A.~Azatov, R.~Contino, C.~S.~Machado and F.~Riva,
  Phys.\ Rev.\ D {\bf 95}, no. 6, 065014 (2017)
  doi:10.1103/PhysRevD.95.065014
  [arXiv:1607.05236 [hep-ph]].

\bibitem{MSXZ2019}
  T.~Ma, J.~Shu, M.L.~Xiao and Y.H.~Zheng,
  in preparation

 

\bibitem{Arkani-Hamed:2017jhn}
  N.~Arkani-Hamed, T.~C.~Huang and Y.~t.~Huang,
  arXiv:1709.04891 [hep-th].


\bibitem{Elvang:2013cua}
  H.~Elvang and Y.~t.~Huang,
  arXiv:1308.1697 [hep-th].

\bibitem{Cheung:2017pzi}
  C.~Cheung,
  arXiv:1708.03872 [hep-ph].

\bibitem{Giudice:2007fh}
  G.~F.~Giudice, C.~Grojean, A.~Pomarol and R.~Rattazzi,
  JHEP {\bf 0706}, 045 (2007)
  doi:10.1088/1126-6708/2007/06/045
  [hep-ph/0703164].

\end{thebibliography}
\end{document}